\documentclass[5p,twocolumn,times,number]{elsarticle}

\usepackage{graphicx}
\usepackage[squaren]{SIunits}

\begin{document}

\begin{frontmatter}

\title{HARPO - A Gaseous TPC for High Angular Resolution Gamma-Ray
 Astronomy and Polarimetry from the MeV to the TeV}

\author[add1]{D.~Bernard\corref{cor}}
\ead{denis.bernard at in2p3.fr}

\address[add1]{LLR, Ecole Polytechnique, CNRS/IN2P3,
91128 Palaiseau, France}

\cortext[cor]{On behalf of the HARPO project, LLR Ecole Polytechnique CNRS/IN2P3 and IRFU/CEA.}

\begin{abstract}
We propose a ``thin'' detector as a
high-angular-precision telescope and polarimeter for cosmic $\gamma$-rays
above the pair-creation threshold.
We have built a demonstrator based on a gaseous TPC.
We are presently characterizing the detector with charged cosmic rays
in the laboratory. Here we present some of its properties.
\end{abstract}


\end{frontmatter}

\section{Introduction}

Understanding the mechanisms responsible for producing high-energy
non-thermal radiation in extreme astrophysical sources, such as AGN
and pulsars, relies on having a complete picture of the emission
across the EM spectrum.

In practice though, huge sensitivity gaps exist, in particular
 \cite{Schoenfelder}
between the domain of high sensitivity of $\gamma$-ray telescopes
based of Compton scattering (mainly sub-MeV) and those based on
$e^+e^-$-pair creation (mainly above 100 MeV).
Several projects exist to improve the sensitivity of Compton telescopes
at the high end of their energy range.

For $e^+e^-$-pair telescopes, the main obstacle is the degradation of
the angular resolution at low energy due to multiple scattering of the
conversion electrons : event selection and background rejection become
increasingly difficult\footnote{Compare, e.g. the effective area after
 successive steps in event reconstruction/selection in Fig. 14 of
 Ref. \cite{Collaboration:2012kc}.}, 
which ultimately limits the sensitivity.

Furthermore, none of the previous (COS-B, EGRET) or present
(FERMI/LAT) $e^+e^-$-pair telescopes have (had) any significant
sensitivity to the polarisation of the incoming photon, a diagnostic
that is a powerful tool in the radio, optical and X-ray bands and
that is missing for $\gamma$-rays.
Gamma-rays are emitted by cosmic sources in a variety of non-thermal
processes, some of which such as synchrotron radiation or inverse
Compton scattering provide linearly polarized radiation to some
extent, while others, such as nuclear interactions, end up producing
non polarized photons.

Most Compton telescope projects include polarimetry, but the
polarization asymmetry of Compton scattering decreases as $1/E$ for $E
\gg m_e c ^2$, so that 
Compton polarimetry becomes inefficient above a couple of MeV
 \cite{McConnell}.

\section{Thick vs thin detector}

The
FERMI/LAT, the $\gamma$-ray mission presently in space, uses a W/Si
converter/tracker followed by a CsI electromagnetic calorimeter : a
thick technology for which the effective area is the product of the
geometrical area by the efficiency, $A_{eff} = S \times \epsilon$.
Over most of the energy range (up to 10 GeV), the angular resolution is dominated by
multiple scattering in the tungsten slabs.

In a thin detector \cite{Bernard:2012em};
\begin{itemize}
\item 
Only a fraction of the
incoming photons are converted, and effective area becomes the product
of the cross-section of the conversion process of interest and the
sensitive mass, $A_{eff} = \sigma \times M$. 

For an argon-based TPC the effective area becomes larger than 
$1 \meter^2 / \ton$ above a couple of MeV (Fig. \ref{fig:Aeff}), while
the effective area of the three-ton FERMI/LAT plateaus at
$\approx 1 \meter^2$ only above 1 GeV \cite{Collaboration:2012kc}.

\begin{figure}[htb]
\begin{center} 
 \includegraphics[width=0.97\linewidth]{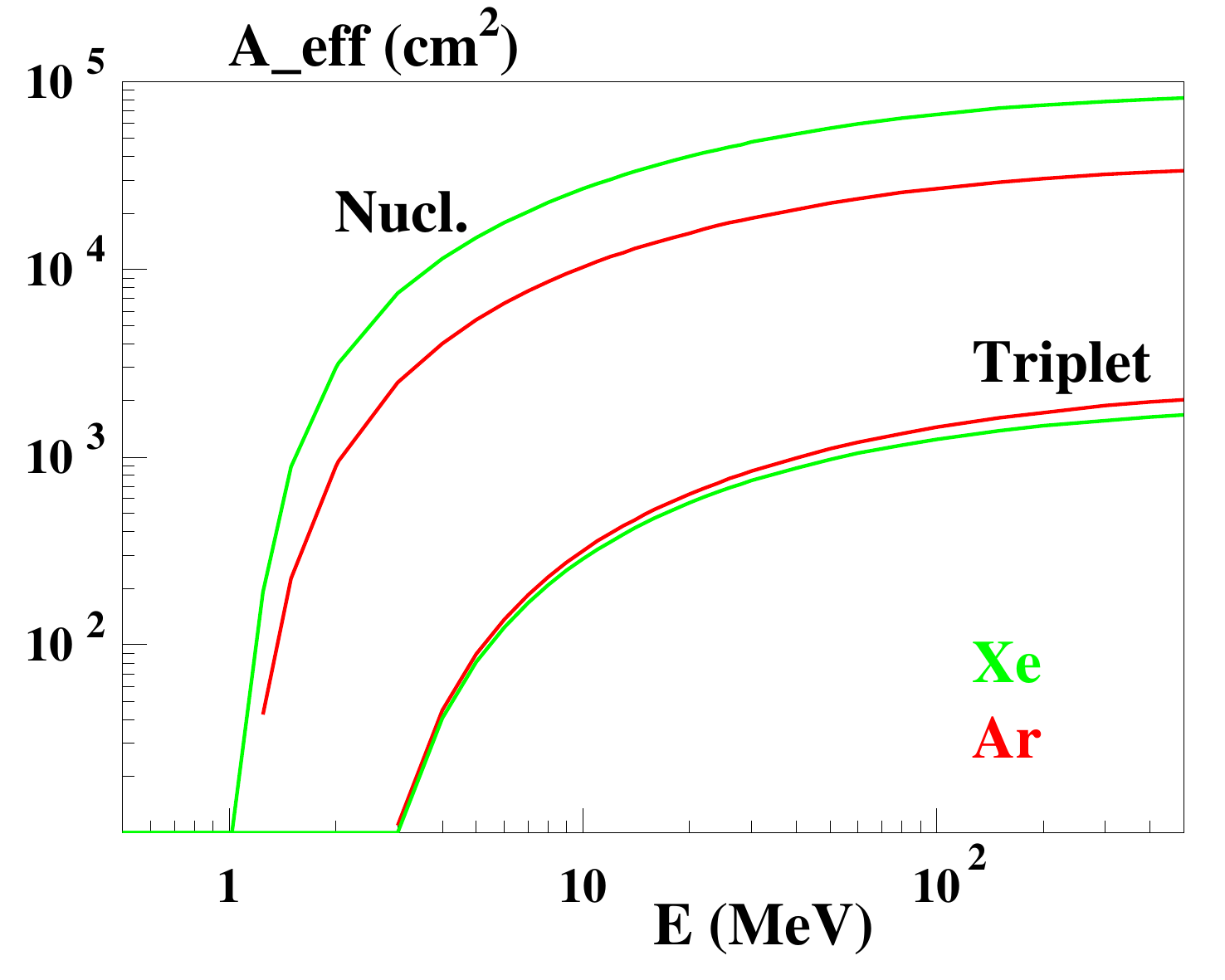}
\caption{\label{fig:Aeff} 
Effective area per ton of sensitive material as a function of the
energy of the incoming photon, for nuclear and triplet pair conversion
(from Ref. \cite{nist}).}
\end{center} 
\end{figure}

\begin{figure}[htb]
\begin{center} 
 \includegraphics[width=0.97\linewidth]{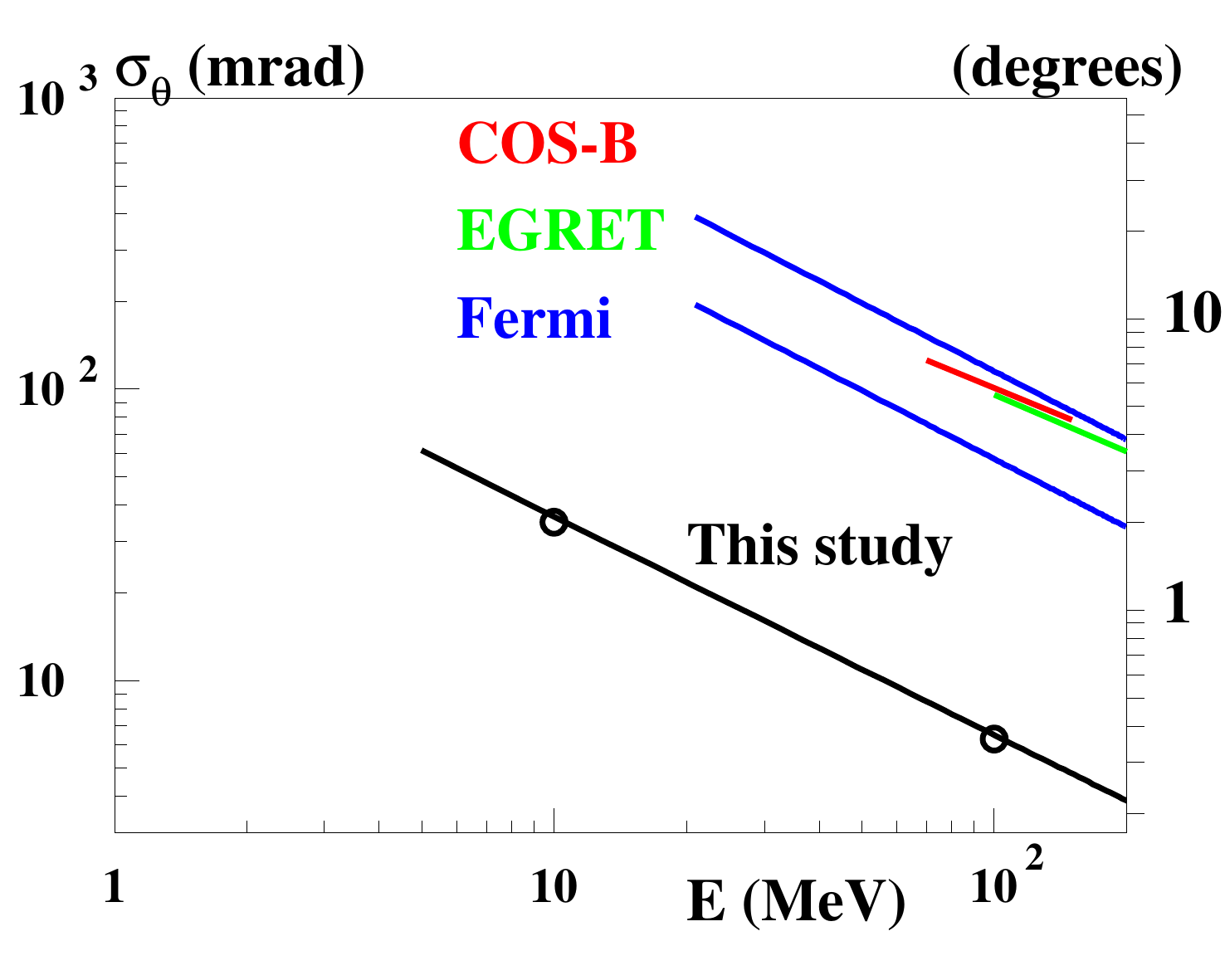}
\caption{\label{fig:resolution:angulaire}
RMS angular resolution as a function of the energy of the incoming photon
for a 5 bar TPC, compared to present/past missions (68\% containment angle).
The Fermi curves (``front'' and ``back'' events) are from the
parametrization of Ref. \cite{Collaboration:2012kc}.}
\end{center} 
\end{figure}

The energy measurement that is obtained from multiple measurements of
multiple scattering of the tracks in the detector at low energy must
be complemented by an additionnal system at higher energy, which will
be chosen to be a ``thin'' system too, so as not to overrun the mass
budget, e.g. using a magnetic spectrometer.

\item 
The angular resolution is improved by one order of magnitude
(figure \ref{fig:resolution:angulaire}), and therefore the
background rejection factor for point-like sources by two orders of
magnitude.

\item Polarimetry is performed by analysis of the distribution of
 the azimutal angle of the recoil electron in triplet conversion,
 i.e. the conversion of the incoming photon in the electric field of
 an electron of the detector, $\gamma e^- \rightarrow e^+e^- +e^-$.
The magnitude of the recoil momentum is typically of the order of 
$\mega\electronvolt/c$, the detection of which requires a low density
medium.
\end{itemize}

A time projection chamber (TPC) i.e., an homogeneous, 3D finely
instrumented medium is particularly well suited for such a detection.

\section{The demonstrator}

We have built a demonstrator 
 consisting of a 5-bar
 argon-based cubic TPC, with a size of 30 cm, and with a pitch,
 sampling frequency and diffusion-induced resolution of about 1 mm.
The amplification is performed with a ``bulk'' Micromegas mesh
 \cite{bulk}, the signal collected by two orthogonal strip sets, and
 digitized with  chips \cite{AFTER} developed for T2K
 \cite{Abgrall:2010hi}.

After we have characterized its performance as a tracker under
(charged) cosmic rays in the laboratory, we plan to expose it to a beam
of linearly polarized $\gamma$ rays \cite{NewSUBARU,Muramatsu:2012uc}
aiming at:

\begin{itemize}
\item validating the technique ($\gamma$-ray astronomy and polarimetry).
\item obtaining the first measurement of the polarization asymmetries
 in the low energy part of the spectrum (few-MeV -- few 10's MeV),
 where the signal peaks, given the spectra of cosmic sources
 (most often a power law $E^{-\alpha}$, where $\alpha \approx 2.$) and where the
 polarization asymmetry is rapidly increasing, and where the approximations
 used in the theoretical calculations (screening, Born approximation) are to be
 validated\footnote{The only experimental validation, to my
  knowledge, was performed at high energy, and with nuclear pair
  conversion \cite{deJager:2007nf}.}.
\end{itemize}

\section{The detector}

\begin{figure}[htb]
\begin{center} 
 \includegraphics[width=0.97\linewidth]{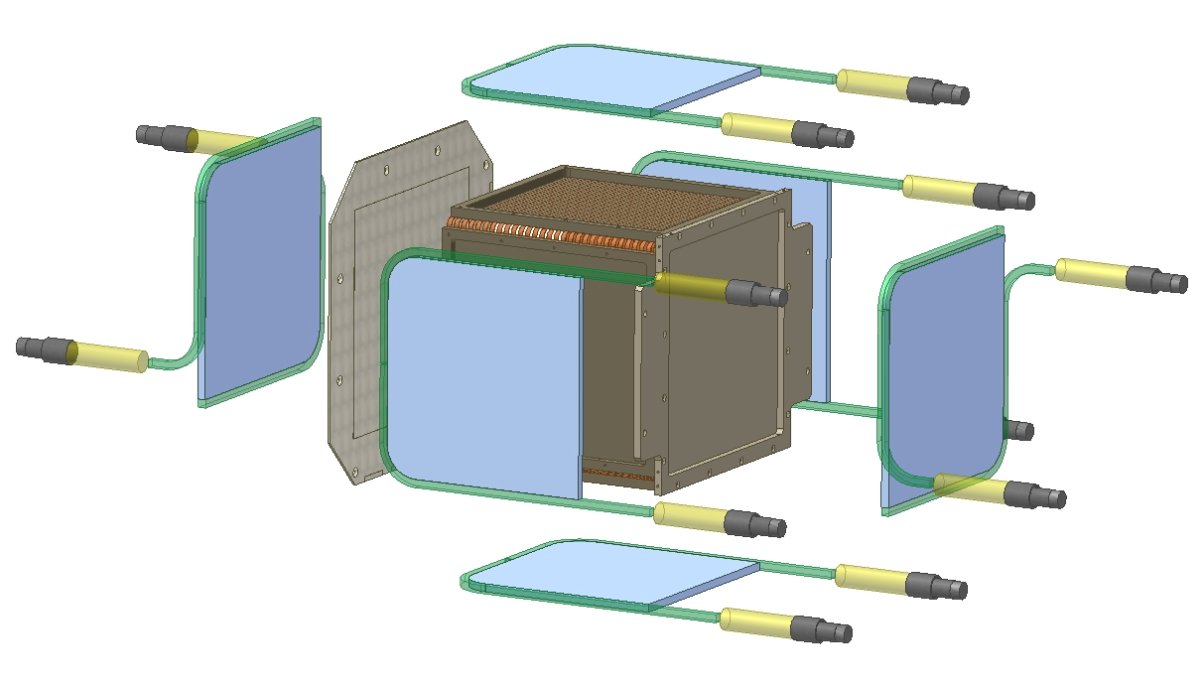}
\caption{\label{fig:layout} Layout of the detector. 
The cubic TPC, with the cathode (right) and Micromegas amplification system (right), surrounded by the trigger scintillator plates.
}
\end{center} 
\end{figure}

The layout of the detector is shown on Fig. \ref{fig:layout}.
\begin{figure}[htb]
\begin{center} 
 \includegraphics[width=0.9\linewidth]{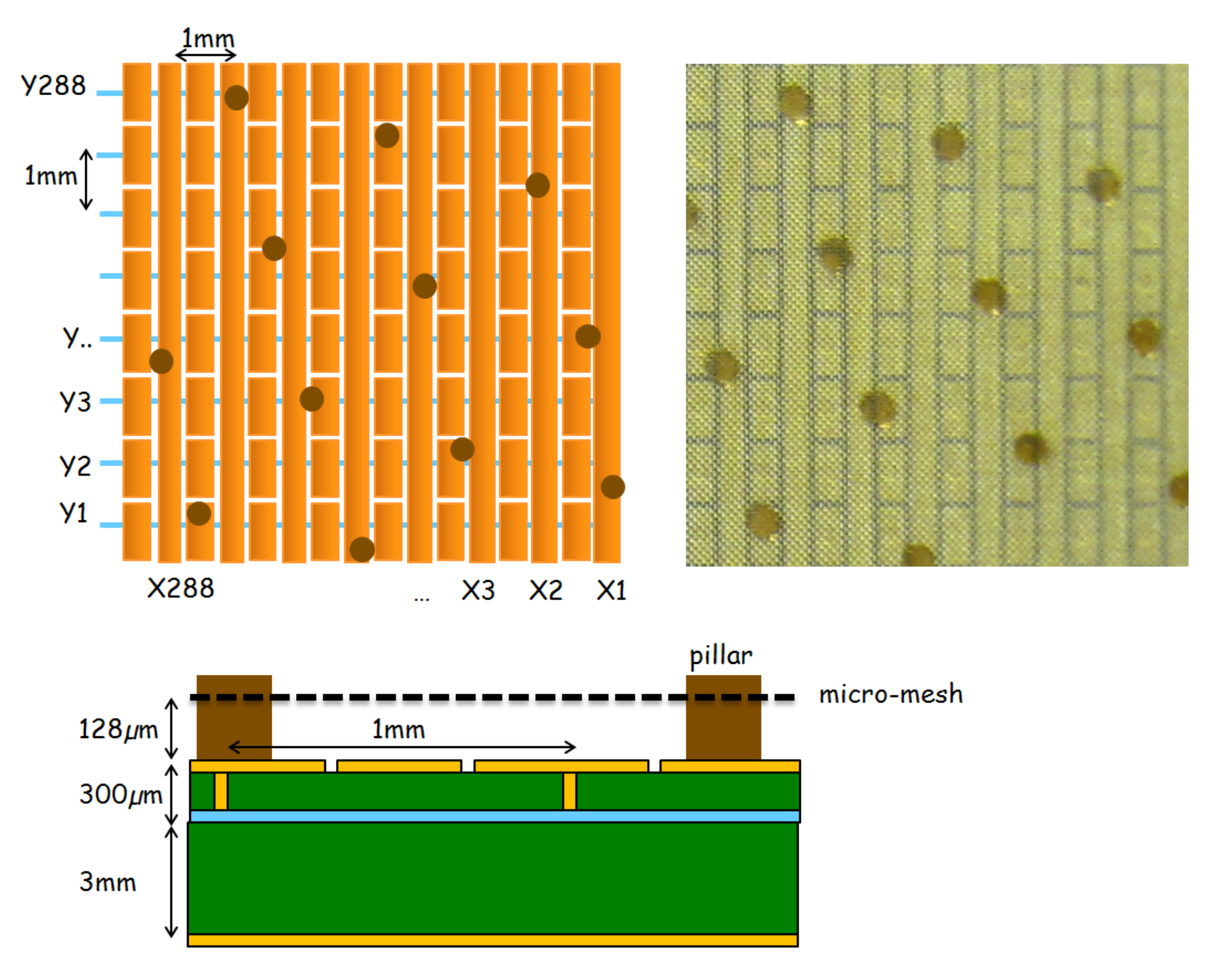}
\caption{\label{fig:mM-and-PCB} 
Top : Layout (left) and photograph (center) of the readout.
Bottom : Schema of the Micromegas.
}
\end{center} 
\end{figure}
The TPC uses the same argon-based mixture (Ar:95 isobutane:2 CF4:3) as
is used by the T2K experiment \cite{Abgrall:2010hi}.

The uniform electric field of the TPC is established by a cubic field
cage, the side of which are 4 PVC plates, thinned down to 1
\milli\meter\ onto which a kapton foil is glued.
The 0.1 \milli\meter\ thick foil is covered by 60 Cu 3mm-wide strips
at a pitch of 5mm, with Cu thickness of 35 \micro\meter.
A resistance divider is built with a set of pairs of resistances of 10
\mega\ohm, which are paired to an homogeneity of about $3. \times 10^{-4}$.

After drift, the ionisation electrons are amplified by a ``bulk''
Micromegas with a gap of 128 \micro\meter, and collected on a PCB
segmented in two series of orthogonal strips, one of which is made of
 regular copper strips, the other of rectangular pads which
are connected by vias to strips lying within the PCB
(Fig. \ref{fig:mM-and-PCB}).

The signal is digitized by an AFTER chip \cite{AFTER}, with 72
channels and a range of 12 bits, a shaping time of 100 ns and a
maximum sampling rate of 50 \mega\hertz, with a maximal number of time
bins of 511.
For each transverse direction ($x, y$), the TPC end plate is
instrumented on a width of 288 \milli\meter, read by four chips
located by one front-end card (FEC) \cite{AFTER}.
The electric field is parallel to $\vec{z}$, so that the third
coordinate of a track point in the detector is obtained from the drift
time, $z = v_{drift} \times (t - t_0)$.

The trigger is built from the signals from 1 \centi\meter\ thick scintillator
 plates \cite{Eljen} which surround the TPC
(Fig. \ref{fig:layout}).
The scintillation light exiting from the edges of the plates are
wavelength shifted in WLS bars. 
The part of the light that propagates along the bar exits from the
pressure vessel and is read by twelve ETL 9125FLB photomultiplier tubes
\cite{Adam:2004fq}.

Data taking can be performed either reading the whole data (``raw'')
or after zero suppression (``Zsuppr'') has been applied in the FPGA of the
AFTER chip, in which case the digitization time lasts for about 5.5
\milli\second.

Data reconstruction includes thresholding, longitudinal (along one
channel) and transverse (at a given bin time) clustering, track
reconstruction with a combinatorial Hough method, and matching of the
two ($x$ and $y$) views.

\begin{figure}[htb]
\begin{center} 
 \includegraphics[width=0.97\linewidth]{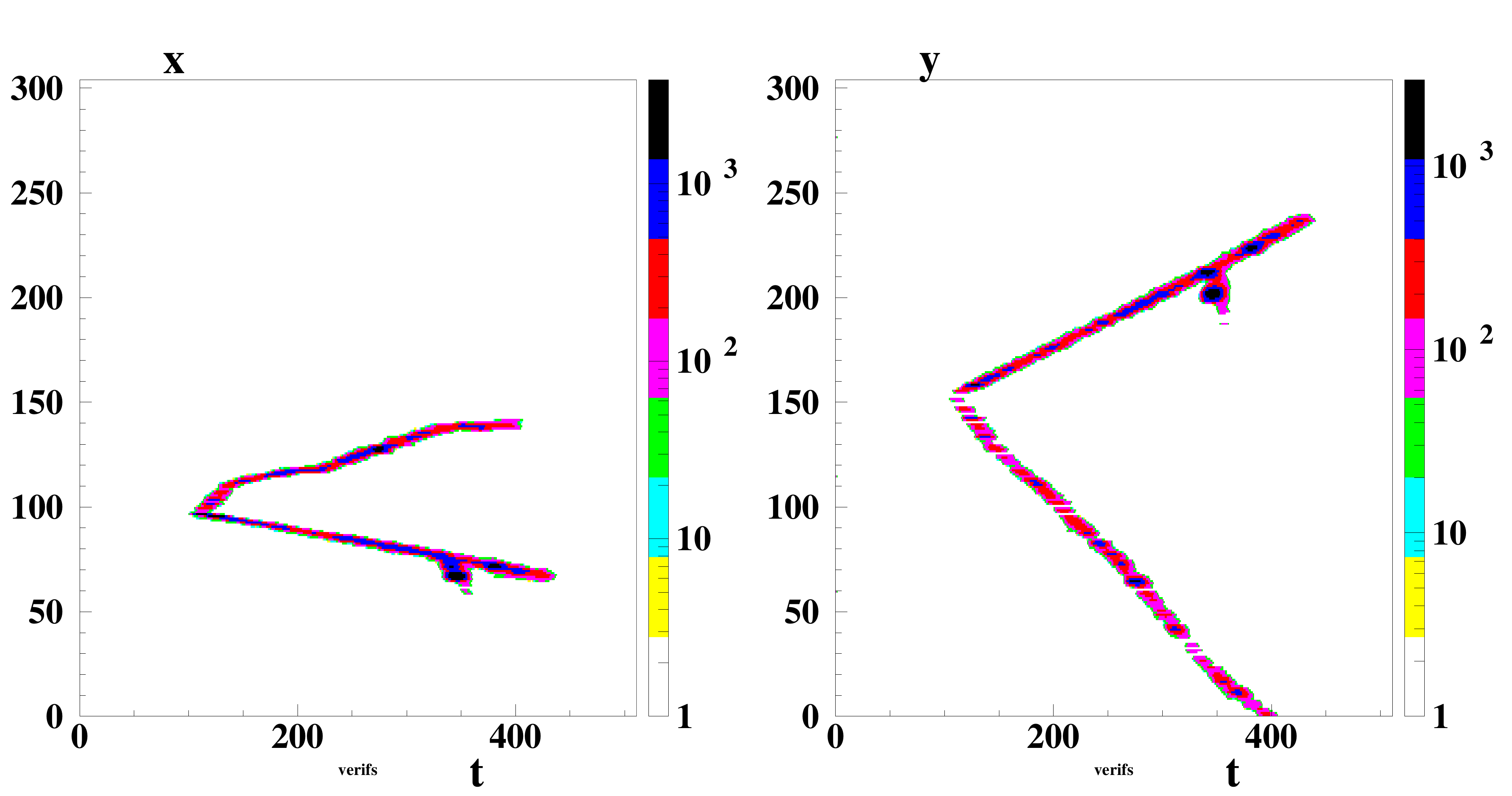}
\caption{\label{fig:oneevent} 
The two views ($x, t$ and $y, t$) of an event (ADC counts).
Units are channel number ($x, y$), time bin ($t$). 
}
\end{center} 
\end{figure}

In cosmic-ray tests most events consist of a simple charged track.
Track $x - y$ matching is most easily explained with a multi-track
event, such as that shown on Fig. \ref{fig:oneevent}.

The delay of the DAQ was set so that the drift starts at time
bin $t=100$, so we most likely see here the interaction of a cosmic
muon (the hard track) with the upper scintillator, producing a delta
ray. A softer delta ray is also visible.

\begin{figure}[htb]
\begin{center} 
 \includegraphics[width=0.97\linewidth]{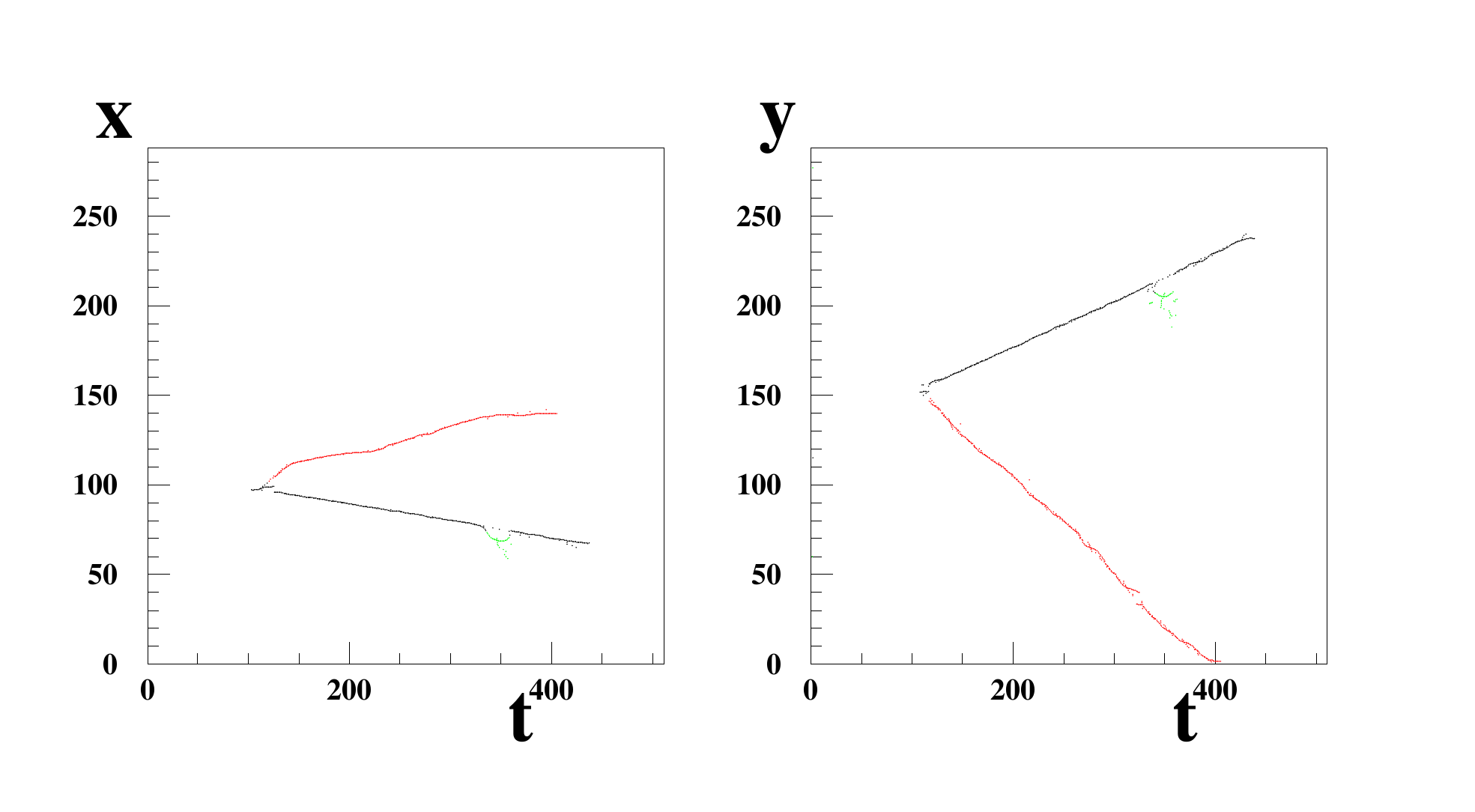}

 \includegraphics[width=0.97\linewidth]{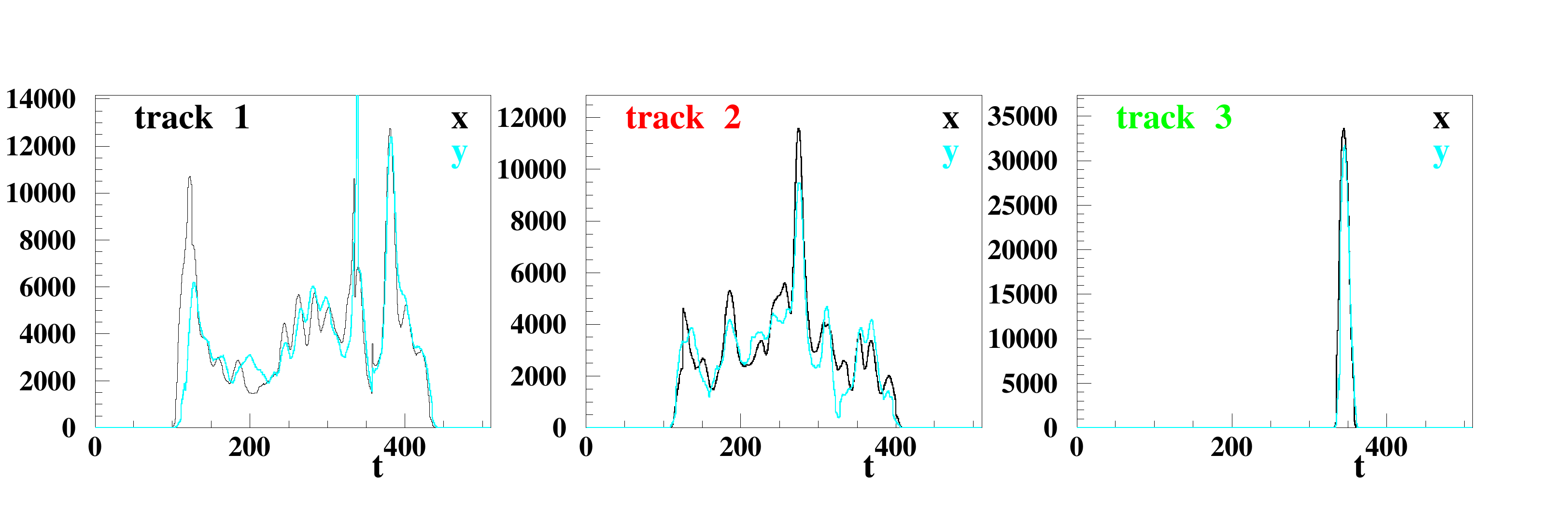}
\caption{\label{fig:xymatching} 
Up : 
The event shown on Fig. \ref{fig:oneevent} after clustering and track reconstruction.
Down : comparison of the $x$ and $y$ time profiles of the three tracks. 
Units are channel number ($x, y$), time bin ($t$). 
}
\end{center} 
\end{figure}

Track $x - y$ matching is performed, comparing the time distributions
of the signal for $x$ and $y$ track. 
Combinations with the smallest $\chi^2$ are chosen 
(Fig. \ref{fig:xymatching}).

\section{Performance}

\begin{figure}[htb]
\begin{center} 
 \includegraphics[width=0.6\linewidth]{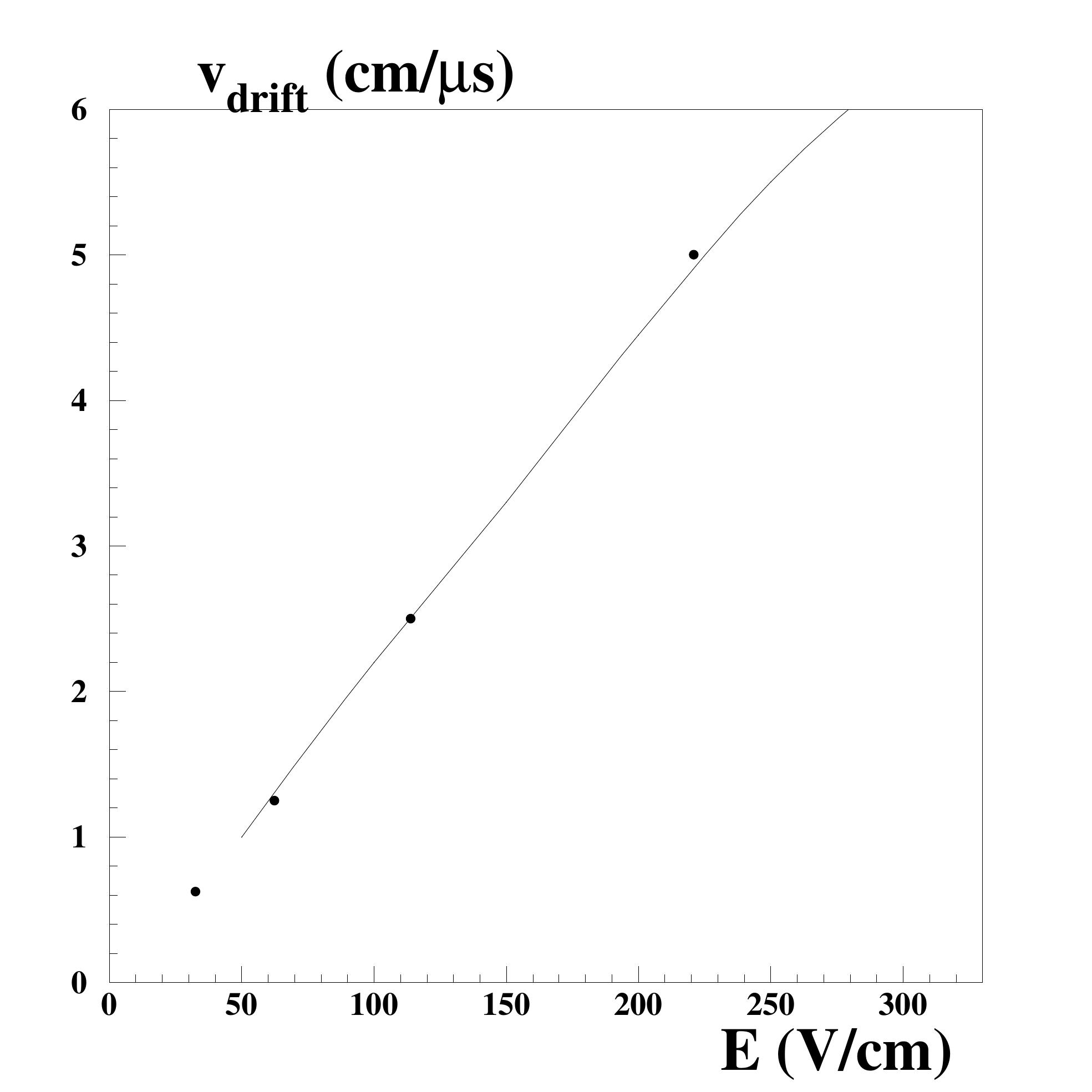}
\caption{\label{fig:drift-velocity} 
Variation of the drift
velocity $v_{drift}$ with the applied electric field, at a gas pressure $P=2 \bbar$. Dots : measurements. Curve : Garfield simulation 
\cite{Veenhof:1998tt}. 
}
\end{center} 
\end{figure}

We are presently characterizing the performance of the detector with
(charged) cosmic rays in the laboratory.
The gas vessel was evacuated to $10^{-4}\milli\bbar$ and then filled
with gas from a pre-mixed bottle.
Data were collected with the axis of the detector in the vertical
position (with the micromegas at the top) : the calibration of the starting time $t_0$ and drift
velocity $v_{drift}$ of the TPC was simply provided by through tracks
(Fig. \ref{fig:drift-velocity}).

\begin{figure}[htb]
\begin{center} 
 \includegraphics[width=0.49\linewidth]{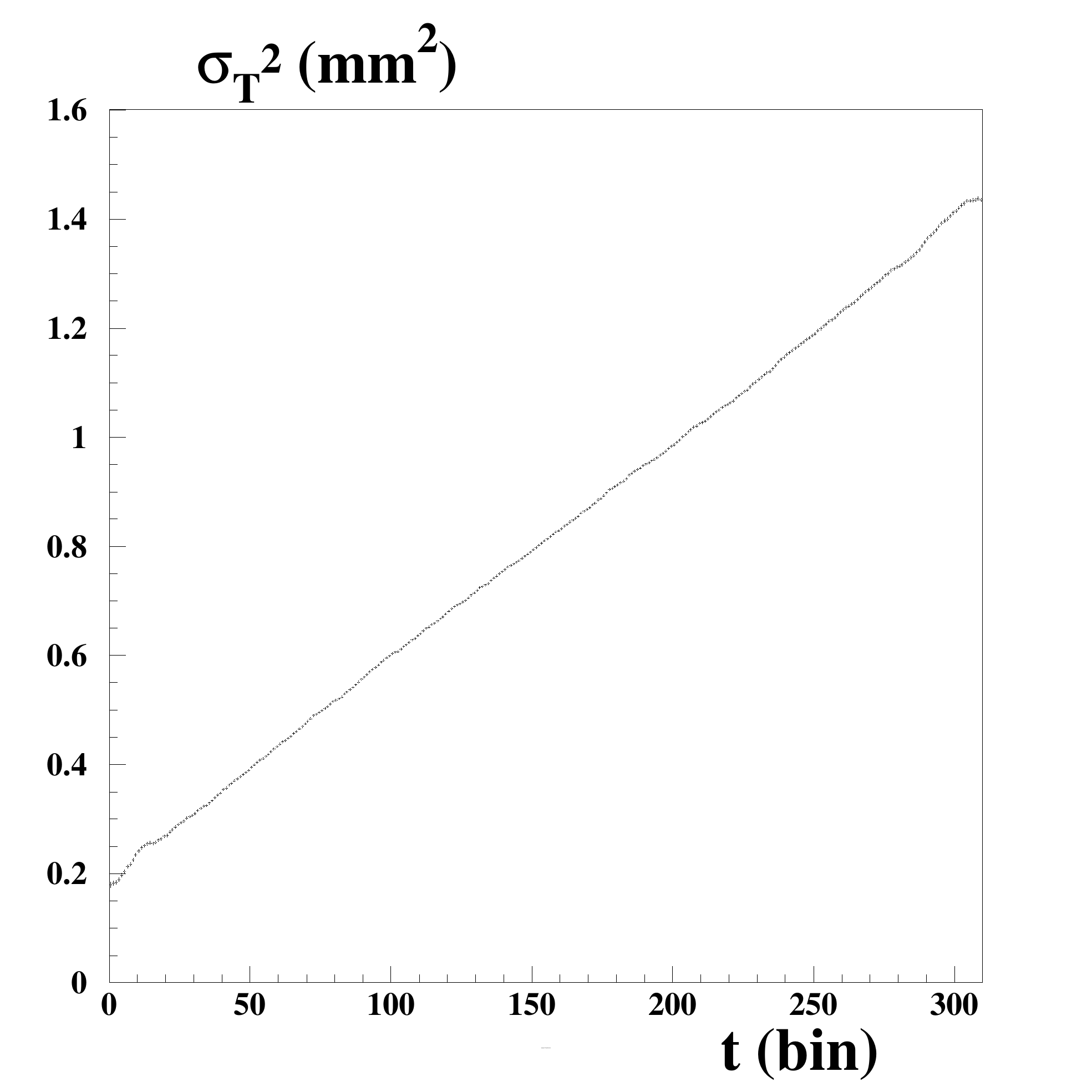}
 \includegraphics[width=0.49\linewidth]{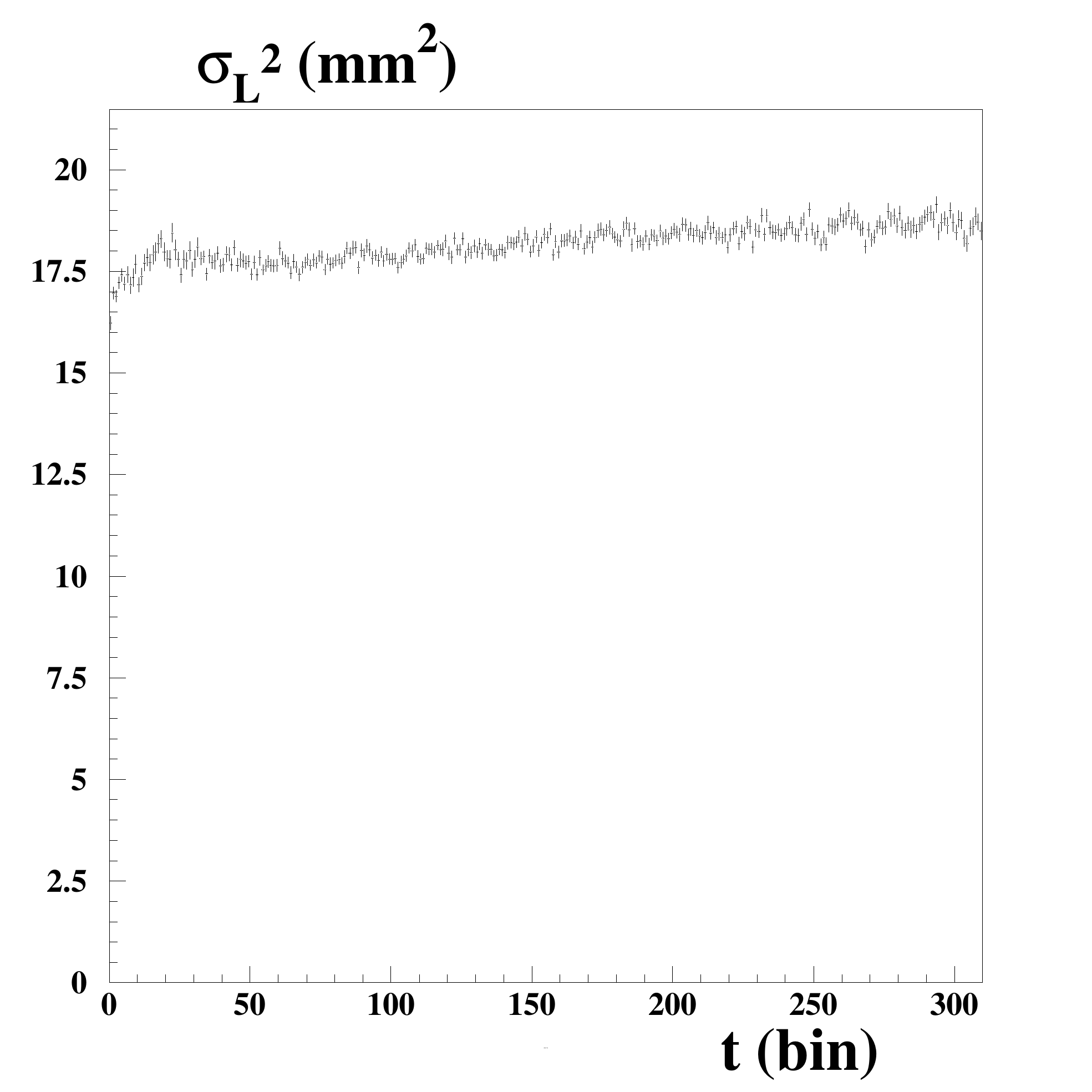}
\caption{\label{fig:diffusion} 
Variation of the squared RMS cluster size 
in the transverse (left) and longitudinal (right) direction, as a function of drift time ($P = 2 \bbar$, $E_{drift} = 234 \volt/\centi \meter$).
}
\end{center} 
\end{figure}

The study of the variation of the squared RMS cluster size as a
function of drift time can provide an estimate of the diffusion of the
electron cloud during the drift.
In the transverse direction, from the linear variation of $\sigma_T^2$
with $t$ for close-to-vertical tracks ($|\theta|< 0.1 \radian$),
(Fig. \ref{fig:diffusion} left), we obtain a transverse diffusion
coefficient of $224 \micro\meter/\sqrt{\centi\meter}$, compatible with
the Garfield simulation.
In the longitudinal direction (right), the shaping of the signal is the
dominating effect.

\begin{figure}[htb]
\begin{center} 
 \includegraphics[width=0.9\linewidth]{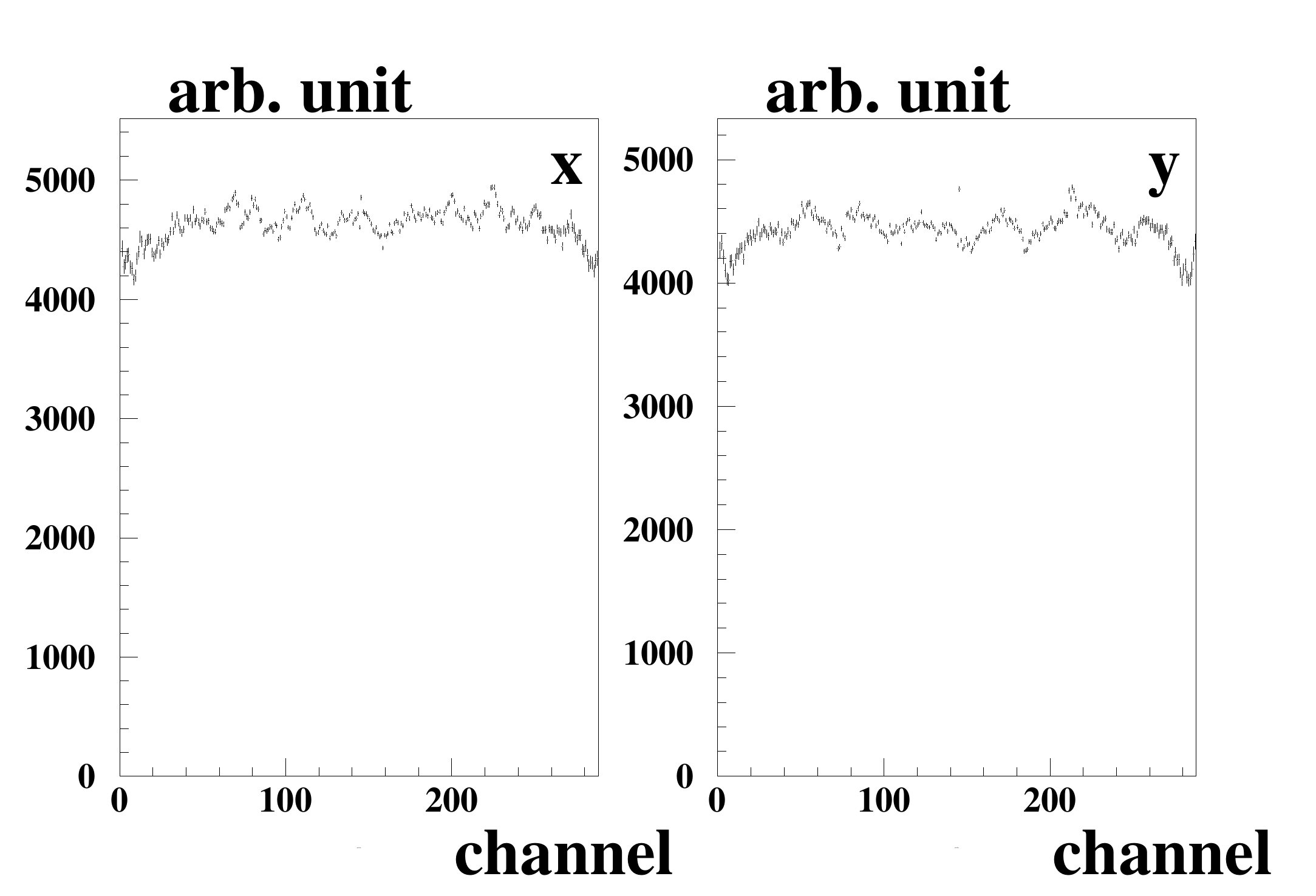}
\caption{\label{fig:homogeneity} 
Map of the gain homogeneity. Left : $x$; Right : $y$.
}
\end{center} 
\end{figure}

From approximately straight, $x/y$ matched tracks, we obtain a map of the
gain homogeneity by dividing the distributions of the position of
their clusters weighted by the cluster total measured charge, by the
unweighted distribution (Fig. \ref{fig:homogeneity}).

\begin{figure}[htb]
\begin{center} 
 \includegraphics[width=0.38\linewidth]{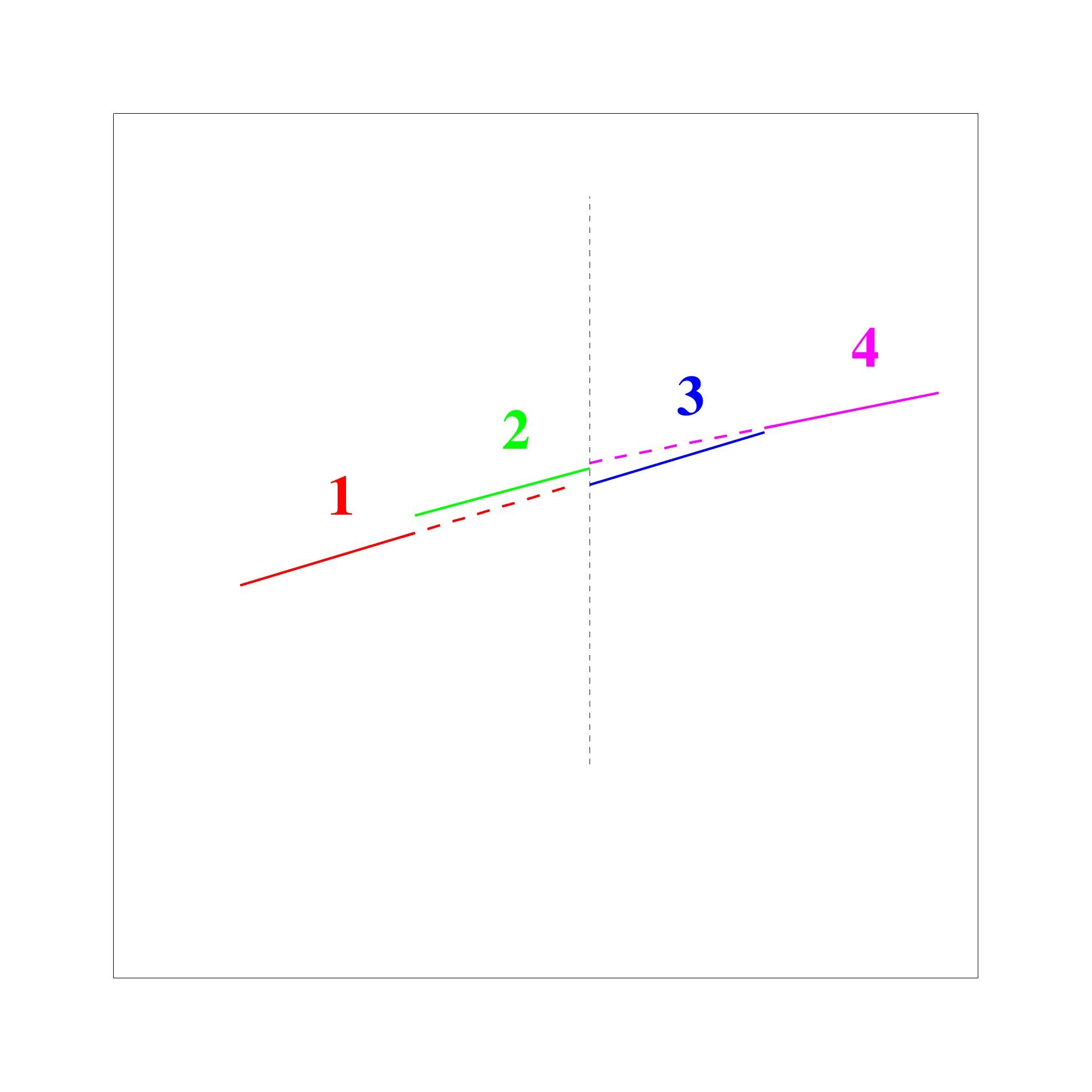}
 \includegraphics[width=0.6\linewidth]{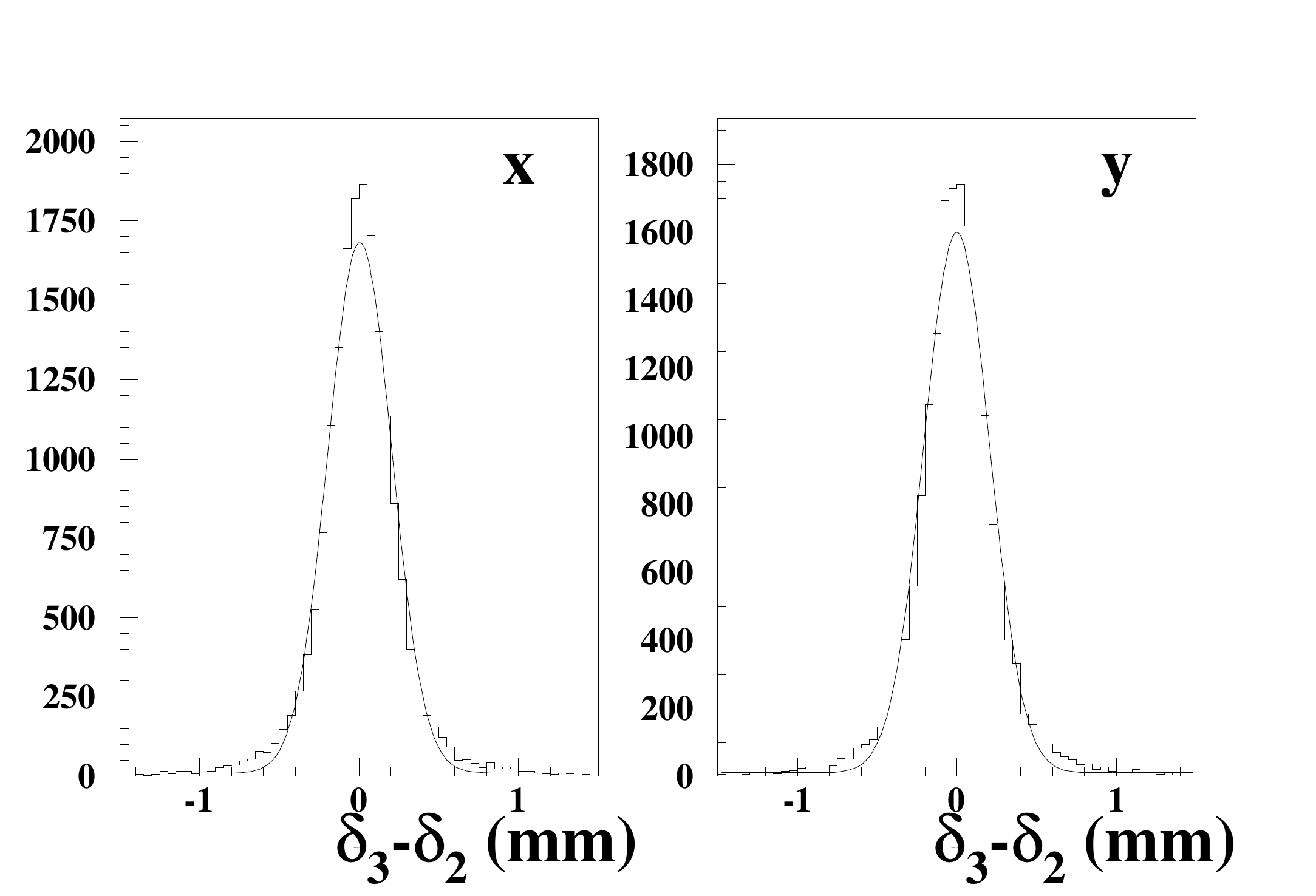}
\caption{\label{fig:resolution} 
Left : Schema of the 4-segments method. Right : Distribution of the
difference in the transverse direction of the extrapolation of segment
2 and 3 to the center of the track. }
\end{center} 
\end{figure}

Finally, an estimate of the spatial resolution is obtained using a
four-segment method.
After a loose preselection of $x/y$ matched tracks, the track is split
into four segments of equal length.
Straight tracks are then selected by applying a cut on the extrapolation
at the track middle from the two external segments 1 and 4.
This does not bias the two internal segments, 2 and 3, the
extrapolations of which are compared.
We obtain a resolution per segment of $200/\sqrt{2} = 140 \micro\meter$ in the
transverse plane ($x$ or $y$, 
Fig. \ref{fig:resolution}).

\section{Conclusion}

We have built a demonstrator for a TPC-based high-angular-resolution
and polarization-sensitive telescope for $\gamma$ rays above the pair
creation threshold.
Aspects of its performance for tracking, obtained from cosmic-ray tests in the
laboratory, have been presented.
We expect to expose this detector to beams of polarized $\gamma$ ray,
to characterize its properties as a $\gamma$-ray telescope and
polarimeter, and to perform the first measurement of the polarization
asymmetry for triplet conversion at low energy.

\bibliographystyle{elsarticle-num}

\end{document}